\documentclass[12pt]{article}
\usepackage{amsmath,graphicx,color,setspace}
\usepackage{ametsoc,subfigure}
\begin{document}

\title{Destabilization of the thermohaline circulation by transient perturbations to the hydrological cycle}

\author{Valerio Lucarini \thanks{lucarini@alum.mit.edu. Please address correspondence to: Valerio Lucarini, Via Palestro 7, 50123 Firenze, Italy. Tel: +393488814008.} \\ Dipartimento di Matematica ed Informatica,
\\Universit\'{a} di Camerino\\ Via Madonna delle Carceri, 62032
Camerino (MC), Italy  \\ \\ Sandro Calmanti and Vincenzo
Artale\\ENEA-CLIM-MOD\\ Via Anguillarese 301, 00060 S. Maria di
Galeria (Roma), Italy }

\maketitle
%\subjclass{Primary 47A15; Secondary 46A32, 47D20}

\newpage

\begin{abstract}
We reconsider the problem of the stability of the thermohaline
circulation as described by a two-dimensional Boussinesq model
with mixed boundary conditions. We determine how the stability
properties of the system depend on the intensity of the
hydrological cycle. We define a two-dimensional parameters' space
descriptive of the hydrology of the system and determine, by
considering suitable quasi-static perturbations, a bounded region
where multiple equilibria of the system are realized. We then
focus on how the response of the system to finite-amplitude
surface freshwater forcings depends on their rate of increase. We
show that it is possible to define a robust separation between
slow and fast regimes of forcing. Such separation is obtained by
singling out an estimate of the critical growth rate for the
anomalous forcing, which can be related to the characteristic
advective time scale of the system.
%
%We present a complete study of the stability of the thermohaline
%circulation using a simplified 2D Boussinesq uncoupled model
%descriptive of the Atlantic ocean. We adopt restoring boundary
%conditions for the temperature and flux boundary conditions for
%the salinity variables. We focus on how the stability properties
%of the system depend on the intensity of the hydrological cycle.
%We consider the 2D parameters' space of the net freshwater fluxes
%into the northern and southern high-latitudes and define, by
%considering suitable quasi-static perturbations to the hydrology,
%a bounded region where multiple equilibria of the system are
%realized. We then consider general time-dependent perturbations to
%the hydrological cycle and we obtain that the rate of change of
%the forcing may dramatically affect the stability of the THC. More
%fundamentally, we show that it is possible to separate robustly
%slow and fast regimes of forcing by singling out a critical rate
%of change of the forcing, which can be related to an estimate of
%the characteristic advective time scale of the system.
\end{abstract}
\newpage
%%% ----------------------------------------------------------------------
%%% ----------------------------------------------------------------------

%%%%%%%%%%%%%%%%%%%%%%%%%%%%%
%%%%%%%%%%%%%%%%%%%%%%%%%%%%%
\section{Introduction}
%%%%%%%%%%%%%%%%%%%%%%%%%%%%%
%%%%%%%%%%%%%%%%%%%%%%%%%%%%%

The thermohaline circulation (THC) plays a major role in the
global circulation of the oceans as pictured by the conveyor belt
scheme \citep{WeaverandHughes1992, Stocker2001}. The currently
accepted picture is that the meridional overturning and the
associated heat and freshwater transports are energetically
sustained by the action of winds and tides, controlling turbulent
mixing in the interior of the ocean
\citep{Munk1998,Rahmstorf2003,WunschandFerrari2004}. However, for
climatic pourposes, many authors have succesfully assumed a
dependence of the strength of the THC from the meridional
gradients in the  buoyancy of the water masses
\citep{Stommel1961,WeaverandHughes1992,
Tzipermanetal.1994,Marotzke1996,Rahmstorf1996,Gnanadesikan1999,
Stockeretal.2001}.

The present day THC of the Atlantic Ocean is characterized by a
strongly asymmetric structure. Deep convection is observed at high
latitudes in the northern hemisphere. The water masses formed in
the northern regions can be followed as they cross the equator and
observed as they connect with the other major basins of the world
ocean \citep{WeaverandHughes1992, Rahmstorf2000, Rahmstorf2002,
Stockeretal.2001}. Idealized and realistic coupled GCM experiments
have shown that such equatorial asymmetry may be a consequence of
the large scale oceanic feedbacks leading to the existence of
multiple equilibria \citep{Bryan1986, ManabeandStouffer1988,
StockerandWright1991, ManabeandStouffer1999a,
MarotzkeandWillebrand1991, HughesandWeaver1994}. The equatorial
asymmetry is also responsible for a large portion of the the
global poleward heat transport \citep{Broecker1994,
RahmstorfandGanopolski1999, Stocker2000, Stockeretal.2001}.
Consequently, large climatic shifts are often associated with
important changes in the large scale oceanic circulation. On a
paleoclimatic perspective, major climatic shifts may be associated
with the complete shutdown of the THC \citep{Broeckeretal.1985,
BoyleandKeigwin1987, Keigwinetal.1994, Rahmstorf1995,
Rahmstorf2002}.

In fact, the THC is sensitive to changes in the climate since the
North Atlantic Deep Water (NADW) formation is affected by
variations in air temperature and in precipitation in the Atlantic
basin \citep{RahmstorfandWillebrand1995, Rahmstorf1996}. With
respect to the present climate change, most GCMs have shown that
the changes in radiative forcing caused by the on-going
modification of the greenhouse gases in the atmosphere could imply
a weakening of the THC. Large increases of the moisture flux
and/or of the surface air temperature in the deep-water formation
regions could inhibit the sinking of the water in the northern
Atlantic \citep{WeaverandHughes1992, ManabeandStouffer1993,
Rahmstorf1997, Rahmstorf1999a, Rahmstorf1999b, Rahmstorf2000,
Wangetal.1999a, Wangetal.1999b}. Moreover, models of different
level of complexity, from box models
\citep{TzipermanandGildor2002,Lucarini2003aa,Lucarini2003bb}, to
EMICs \citep{StockerandSchmittner1997,SchmittnerandStocker1999} to
GCMs \citep{StoufferandManabe1999, ManabeandStouffer1999a,
ManabeandStouffer1999b,ManabeandStouffer2000} have shown how the
rate of increase of forcing may be relevant for determining the
response of the system. In particular
\cite{StockerandSchmittner1997} have performed a systematic
analysis of the stability of the meridional overturning
circulation as a function of the climate sensitivity and of the
rate of CO$_2$ increase. However they made no explicit reference
as to the mechanism driving the response of their coupled model.

In this work we study the THC stability using a simplified 2D
Boussinesq ocean model, which has been presented in
\cite{Artale2002}. Two-dimensional models have been widely adopted
\citep{CessiandYoung1992, Vellinga1996} and have proved their
ability in describing the most relevant feedbacks of the system
\citep{Dijkstra1999}. Moreover, the low computational cost of such
models permits extensive parametric studies. Our work wishes to
bridge both in terms of methodology and results the studies
performed with simplified models with the more physically sensible
analyses performed with EMICs and GCMs. We explicitly analyze what
is the role of the rates of changes of the hydrological forcing in
determining the response of the system. In particular we
determine, for a given initial state and a given rate of increase
of the forcing, which are the thresholds in total change of the
forcing beyond which destabilization of the THC occurs. The
treatment of a wide range of temporal scales for the increase of
the forcing allows us to join on naturally and continuously
\citep{Lucarini2003aa,Lucarini2003bb} the analysis of quasi-static
perturbations, which have been usually addressed with the study of
the bifurcations of the system \citep{Rahmstorf1995,
Rahmstorf1996,
StoneandKrasovskiy1999,Scottetal.1999,Wangetal.1999a,Titzetal.2002a,Titzetal.2002b},
with the study of the effects of very rapid perturbations
\citep{Rahmstorf1996,Scottetal.1999,WiebeandWeaver1999}, which are
usually differently framed.

Our paper is organized as follows. In section \ref{model} we
provide a description of the model we adopt in this study. In
section \ref{stability} we explore the parameters space
descriptive of the hydrology of the system by considering
quasi-static perturbations. We determine which hydrological
patterns are compatible with multiple equilibria and which
hydrological patterns define a unique stationary state. In section
\ref{generalstability} we extend the analysis to time-dependent
perturbations. We analyze the temporal evolution of finite
amplitude modifications of the hydrological cycle that are able to
destabilize the equilibria of this advective system. We propose a
simple relation between an estimate of the critical rate of
increase of forcing, which divides robustly \textit{slow} from
\textit{fast} regimes, with an estimate of the characteristic
advective time scale of the system. In section \ref{concl} we
present our conclusions. In appendix A we present the dependence
of the THC strength on the value of the vertical mixing
coefficient.
%In appendix B we show
%what is the relevance of the meridional temperature gradient for
%the purposes of this study.

%%%%%%%%%%%%%%%%
%%%%%%%%%%%%%%%%
\section{The model}
\label{model}
%%%%%%%%%%%%%%%%
%%%%%%%%%%%%%%%%

We consider the two-dimensional convection equations in the
Boussinesq approximation. The motion is forced by buoyancy
gradients only: gravity is the only external force, while Earth
rotation is not explicitly considered. Buoyancy gradients are
generated in the model by imposing heat and freshwater fluxes at
the top boundary. Such fluxes are assumed to be representative of
the interactions with the overlying atmosphere.

We adopt a linearized equation of state for the sea water:
\begin{equation}
\rho\left(T,S\right)=\rho_0\left(1-\alpha T+ \beta S\right)
\end{equation}
where the values of the coefficients of thermal expansion
$\alpha=8\cdot 10^{-4}$ $^{\circ}C^{-1}$ and haline contraction
$\beta=1.5\cdot 10^{-3}psu^{-1}$ have been chosen in order to
provide a good approximation over a quite large range of salinity
and temperature. The linear approximation is commonly adopted in
conceptual models. However, we note that analyses performed on
simple models show that small nonlinearities may induce
self-sustained oscillations for the THC \citep{Rivin1997}. Such
nonlinearities in the equation of state might be especially
relevant for the high latitude areas, since the well-known
cabbeling effect occurs at low temperatures.

The geometry of the model is descriptive of the Atlantic ocean,
where we assume a depth of $5000m$, an effective east-west
extension of $6000Km$, and a north-south extension of $13600Km$
($120^{\circ}$) and a total volume $V=4.08\times10^{17}m^3$. The
only active boundary of the model is the air-sea interface.
\par We select a relatively coarse uniform resolution with $N_H\times
N_V=64\times16$ grid points, where $N_H$ and $N_V$ refer to the
number of the horizontal and vertical grid points, respectively,
in order to meet the computational requirements needed for
performing a parametric study.

A well known property of the THC system is the existence of
regimes of multiple equilibria. When equatorially symmetric
surface forcing is applied at the surface, the equilibria of the
system fall into three well-distinct classes. One class is
characterized by the presence of two equatorially symmetric
thermally direct cells, where the deep water is formed at high
latitudes. Another class is characterized by equatorially
symmetric salinity driven cells, where deep water is formed in the
equatorial region. Equilibria belonging to these two classes are
observed when either surface thermal or haline buoyancy forcings
are largely dominant, respectively. When the two forcings have
comparable intensity, multiple equilibria regimes - which
constitute a third class - appear. In this case, the equilibria
are characterized by the dominance of one overturning cell. If
also geometry is symmetric with respect to the equator, the two
equilibria have odd parity and map into each other by exchanging
the sign of the latitude.

%%%%%%%%%%%%%%%%

%%%%%%%%%%%%%%%%%%%%%
\subsection{Boundary conditions}
The boundary condition for the sea surface salinity is defined in
terms of the imposed atmospheric freshwater flux $F$ affecting the
surface grid box of volume $v=V/\left(N_H\times N_V\right)$:
\begin{equation}\label{salinity}
\partial_t
S_{i,j=1}=-F_{i}\frac{S_0}{v},
\end{equation}
where we indicate the value of the bulk variable $S$ at the grid
point $\left(i,j\right)$ with $S_{i,j}$ and the value of the
interface variable $F$ at the grid point $\left(i,1\right)$ with
$F_i$. We underline that, in order to simplify the expressions, in
equation (\ref{salinity}) (and in the following ones) we have not
explicitly adopted a discrete notation for the time variable.

We emphasize that in expression (\ref{salinity}) we have neglected
the contribution in terms of mass of the freshwater flux to the
ocean \citep{Marotzke1996}.

\begin{figure}
 \centering
\includegraphics[angle=270, width=1.0\textwidth]{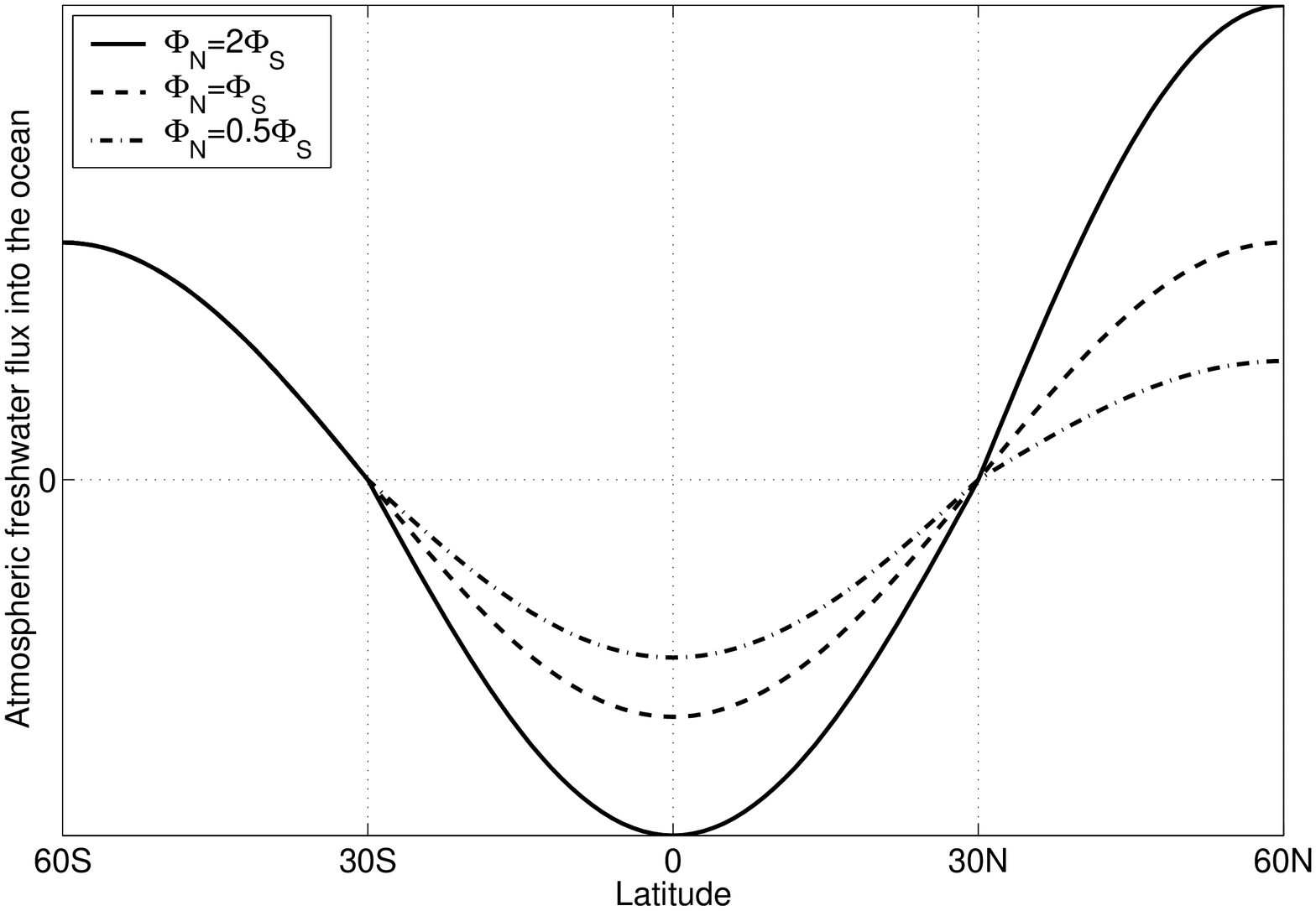}\\
\caption{Surface freshwater flux for three different
configurations of the hydrological cycle.} \label{fluxus}
\end{figure}

We divide the water basin into three distinct regions by using a
suitable analytical expression of the freshwater flux. The
equatorial region (region E) is characterized by a net atmospheric
export of freshwater, while the northern and southern high
latitude regions (regions $N$ and $S$, respectively) are
characterized by a positive atmospheric freshwater budget. We then
consider the following functional form for the surface freshwater
flux:
\begin{equation}\label{fresh}
F_i=\frac{2}{\pi}\Phi_i\cos\left(2\pi i/N_H\right)
\end{equation}
where:
\begin{equation}\label{phi}
\Phi_i=\begin{cases}\Phi_S, \hspace{4mm} i \leq 1/4\hspace{1mm}N_H
\\
\Phi_N, \hspace{4mm} i\geq 3/4\hspace{1mm}N_H
\\
\Phi_E=-1/2\left(\Phi_N+\Phi_S\right), \hspace{4mm}
1/4\hspace{1mm}N_H < i < 3/4\hspace{1mm}N_H
\end{cases}
\end{equation}
The definition of $F_i$ is such that $\Phi_N$ and $\Phi_S$ are
respectively the value of the total net atmospheric freshwater
fluxes into regions $N$ and $S$, while $\Phi_E$ is constrained in
order to have conservation of the salinity of the ocean. This
latter condition is needed to allow the system to reach
equilibrium states. Therefore, the ocean conserves its average
salinity $S_0=35\hspace{1mm}psu$. Since the atmospheric freshwater
budgets for the Atlantic ocean are thought to be positive for the
regions $N$ and $S$ \citep{BaumgartnerandReichel1975}, we consider
the case $\Phi_N,\Phi_S\geq 0$. Three relevant examples of surface
freshwater flux are depicted in figure \ref{fluxus}.

The sea surface temperature is restored to a time-independent
climatological temperature field $\overline{T}_i$ with a newtonian
relaxation law:
\begin{equation}\label{temperature}
\partial_t
T_{i,j=1}=\lambda\left[\overline{T}_i-T_{i,j=1}\right]
\end{equation}
where the constant $\lambda$ describes the efficiency of the
process. Such very simplified ocean-atmosphere coupling
\citep{MarotzkeandStone1995, Marotzke1996} synthetically describes
the combined effects of radiative heating-cooling and of the
atmospheric latent and sensible heat meridional transport. The
climatological temperature profile $\overline{T}_i$ profile is set
with the following equatorially symmetric analytical form:
\begin{equation}\label{tau}
\overline{T}_i=\overline{T}_0-\Delta \overline{T}\cos\left(2\pi i/
N_H \right),
\end{equation}
where $2\Delta \overline{T}$ is the  imposed equator-to-pole
temperature gradient and $\overline{T}_0$ is the average value of
$\overline{T}_i$. In accordance to definition of the $E$, $N$, and
$S$ regions with respect to their atmospheric freshwater budget,
we observe that regions $N$ and $S$ are \textit{cold} at surface,
\textit{i.e.} $\overline{T}_i$ is lower than $\overline{T}_0$,
while the region $E$ is \textit{warm} at surface, \textit{i.e.}
$\overline{T}_i$ is larger than $\overline{T}_0$.

Since in our work we explore the stability of the THC with respect
to the hydrology of the system, we keep fixed  the parameters
determining the restoring temperature profile $\overline{T}_i$. We
set $\overline{T}_0=15^{\circ}C$, since it represents a reasonable
average surface climatological temperature, and we choose $\Delta
\overline{T} \sim 23.5 ^{\circ} C$, which corresponds to forcing
the surface equator-to-pole temperature gradient to be $\sim 47
^{\circ}C$. Such a choice also implies that the average
$\overline{T}$ is $\sim 0^{\circ}C$ in the two high latitude
regions $N$ and $S$ and is $\sim 30^{\circ}C$ in the equatorial
region $E$. Furthermore, following the parameterization proposed
by \cite{Marotzke1996}, we choose the restoring constant $\lambda
\sim (1 y)^{-1}$, which is reasonable for the description of the
temperature relaxation of the uppermost $5000/16\hspace{1mm}m \sim
300\hspace{1mm}m$ of the ocean.
%\par A very relevant parameter in the 2D ocean models is the
%vertical diffusivity $K_V$; we explore the sensitivity of our
%system with respect to its value in appendix A. In our simulations
%we choose the commonly used value $K_V=1 cm^2s^{-1}$.

%%%%%%%%%%%%%%%%%%%%%%%%%%%%%%%%%%%%%%%%
%%%%%%%%%%%%%%%%%%%%%%%%%%%%%%%%%%%%%%%%
\section{Quasi-static hysteresis and multiple equilibria}
\label{stability}
%%%%%%%%%%%%%%%%%%%%%%%%%%%%%%%%%%%%%%%%
%%%%%%%%%%%%%%%%%%%%%%%%%%%%%%%%%%%%%%%%
\subsection{Symmetric case: $\Phi_S=\Phi_N$}
\begin{figure}
\centering
\includegraphics[angle=270, width=1.0\textwidth]{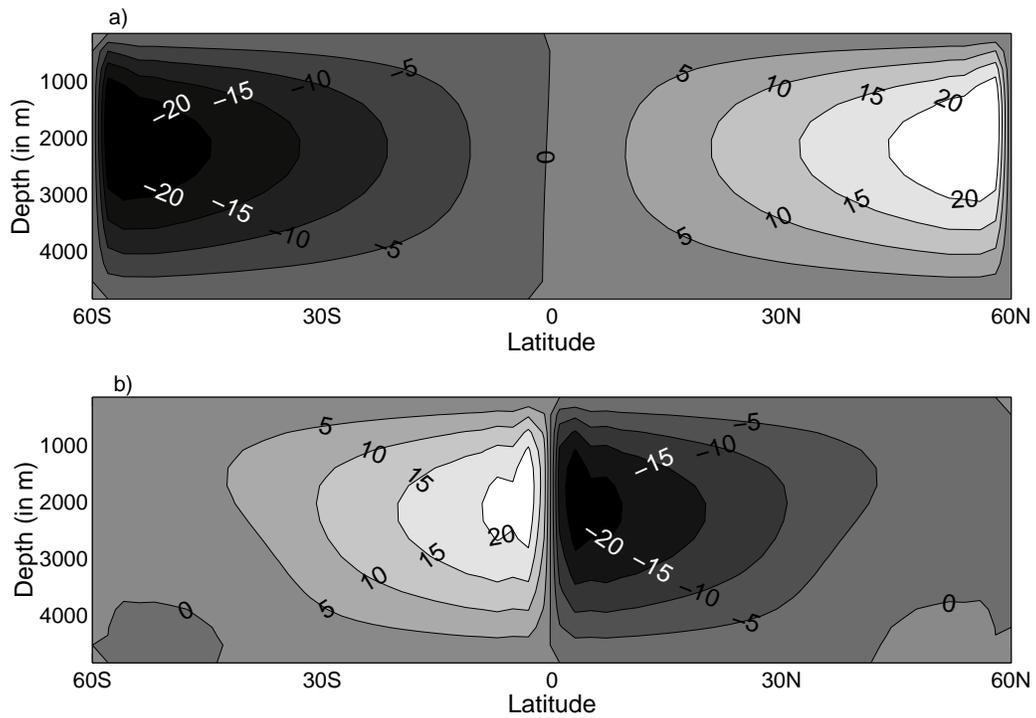}\\
\caption{Circulation patterns obtained in the non-bistable region
for symmetric forcing. Transport is expressed in Sv. a):
$\Phi_S=\Phi_N=0.1\Phi_{av}=0.5\Phi_{inf}$. b) :
$\Phi_S=\Phi_N=2\Phi_{av}=1.1\Phi_{sup}$.}\label{PE}
\end{figure}

\begin{figure} \centering
\includegraphics[angle=270, width=1.0\textwidth]{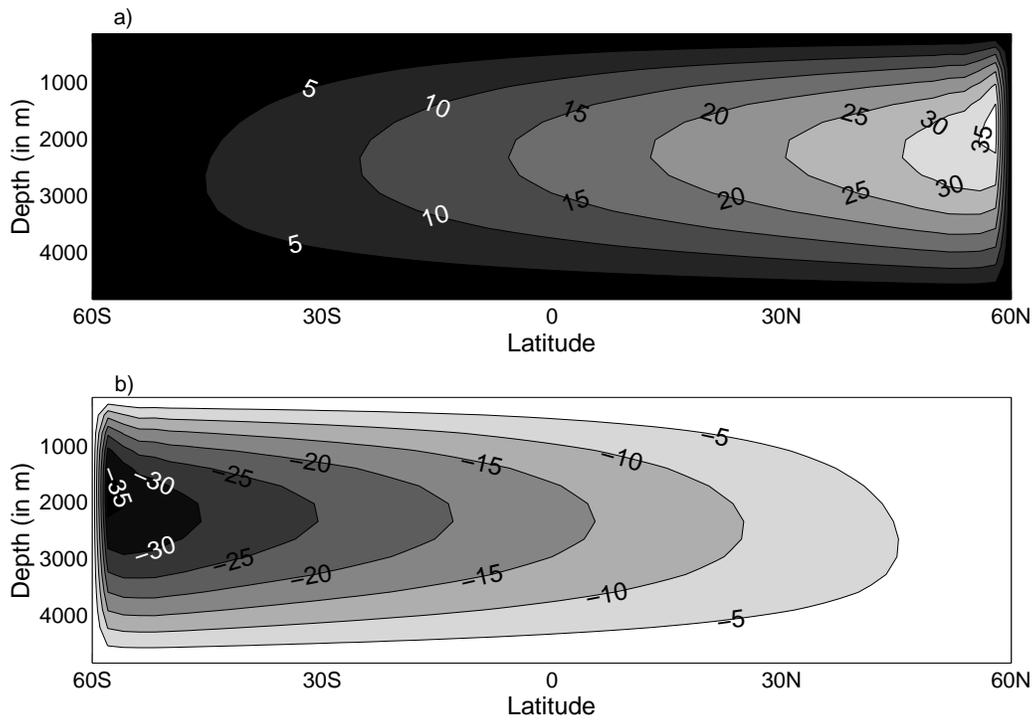}\\
\caption{Circulation patterns obtained in the bistable region for
symmetric forcing with $\Phi_S=\Phi_N=\Phi_{av}$. Transport is
expressed in Sv. a) Northern sinking pattern. b) Southern sinking
patterns.}\label{NS}
\end{figure}

We start by considering the existence of a region of multiple
equilibria in the space of parameters that defines the hydrology
of the system, \textit{i.e.} the $\left(\Phi_S,\Phi_N\right)$
plane. The surface thermal forcing is kept fixed. In the space of
parameters that we are considering, symmetric forcings constitute
the bisectrix of the $\left(\Phi_S,\Phi_N\right)$ plane. The
allowed circulation patterns change when adjusting the parameters
along the bisectrix, as discussed in section \ref{model}. For a
weak hydrological cycle ($\Phi_N=\Phi_S<\Phi_{inf}$), we observe a
stable symmetric circulation with downwelling at high latitudes.
If the hydrological cycle is very strong
($\Phi_N=\Phi_S>\Phi_{sup}$), we obtain a stable symmetric
circulation with downwelling of warm, saline water at the equator.
In the intermediate regimes ($\Phi_{inf}\leq \Phi_N=\Phi_S\leq
\Phi_{sup}$), the system has multiple equilibria. For later
convenience, we define $\Phi_{av}\equiv
0.5\left(\Phi_{inf}+\Phi_{sup}\right)$, which corresponds to an
\textit{average} hydrological cycle. The described equilibria are
depicted in figures \ref{PE}a), \ref{PE}b), \ref{NS}a), and
\ref{NS}b) respectively. The two points
$(\Phi_S=\Phi_{inf},\Phi_N=\Phi_{inf})$ and
$(\Phi_S=\Phi_{sup},\Phi_N=\Phi_{sup})$  are bifurcation points in
the one-dimensional subspace $\Phi_N=\Phi_S$ \citep{Dijkstra1999}.
We consider the northern sinking equilibrium state (note that
northern and southern sinking patterns are equivalent in these
terms) realized for $\Phi_S=\Phi_N=\Phi_{av}$ as the reference
state of the system.

\subsection{General case: $\Phi_S\neq\Phi_N$}

The study of the multiple equilibria states can be extended to the
case of non-symmetric forcings, \textit{i.e.} $\Phi_S \neq
\Phi_N$. We wish to obtain an estimate of the shape of the domain
$\Gamma$ in the $(\Phi_S,\Phi_N)$ plane where the system has
multiple equilibria.

Since, apart from the freshwater flux boundary conditions, the
model is wholly symmetric with respect to the equator, we expect
that any property of the system is invariant for exchange of
$\Phi_N$ and $\Phi_S$, so that we have that $\Gamma$ is \textit{a
priori} symmetric with respect to the bisectrix.

A fundamental property of the region $\Gamma$ is that, if we start
from a point belonging to $\Gamma$ and change quasi-statically
$\Phi_S$ and $\Phi_N$ along a closed path so that the point
$(\Phi_S,\Phi_N)$ remains inside $\Gamma$, we get back to the
initial equilibrium state. Instead, if the closed path crosses the
boundary of $\Gamma$, the initial equilibrium state may not be
recovered at the end of the loop, since the final state depends on
the path.

As a starting position of the loop, we consider a northern sinking
equilibrium corresponding to one of the stable states of the point
$\Phi_{inf} \leq \Phi_S=\Phi_N\leq \Phi_{sup}$ on the bisectrix.
By definition, such point belongs to $\Gamma$. However, the
initial position of the loop - if belonging to $\Gamma$ - is not
relevant in determining the shape of the region of multiple
equilibria.

\begin{figure}[t]
\centering
 \subfigure[Schematization of the $\Phi_N$ time-dependence for the three cases of loop within $\Gamma$, loop crossing the border of $\Gamma$, and loop reaching the border of $\Gamma$.]
   {\includegraphics[angle=270,width=0.7\textwidth]{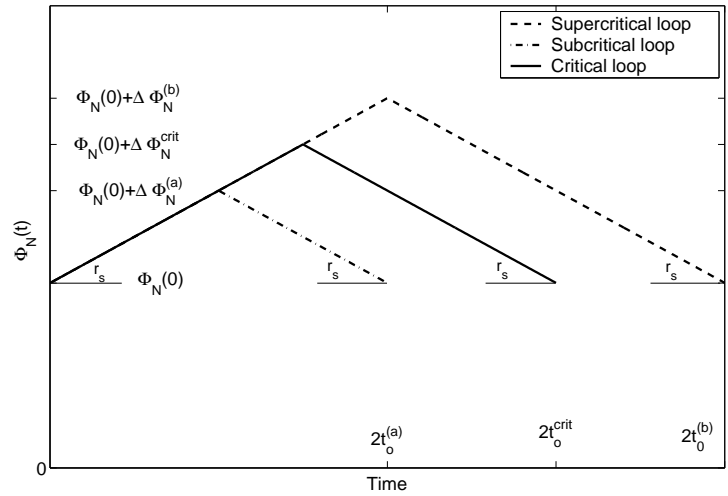}}\\
 \subfigure[Schematization of the $\Phi_N$ time-dependence for three cases of transient forcings.]
   {\includegraphics[angle=270,width=0.7\textwidth]{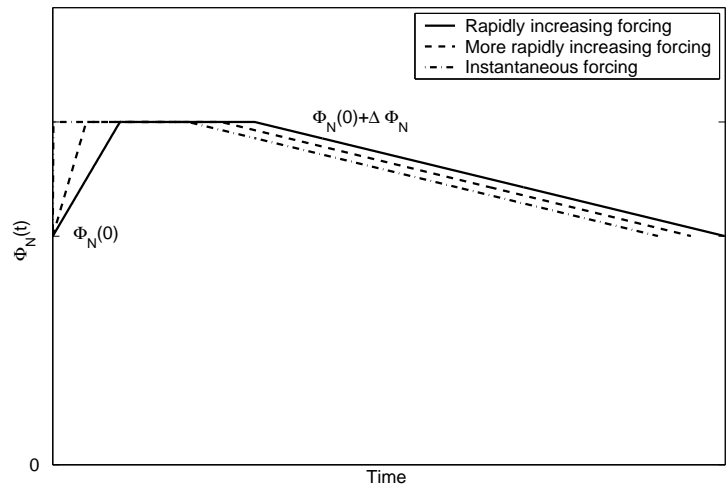}}
 \caption{Quasi-static and transient changes in the value of the parameter $\Phi_N$}\label{changes}
 \end{figure}

We increase the value of $\Phi_N$ at a given slow constant rate
$r_s$ over a time $t_0$ and then decrease it back to the initial
value at the rate $-r_s$. By slow rate we mean that $r_s \ll
\Phi_{av} / \tau$; we select $r_s=\Phi_{av} / 100 \tau$. If the
initial state is not recovered at the end of the integration, we
deduce that the path has crossed the boundary of $\Gamma$. By
bisection, we can determine the critical value $t_0^{crit}$, which
determines $\left(\Phi_S,\Phi_N+r_s\cdot t_0^{crit}=\Phi_N+\Delta
\Phi_N^{crit}\right)$ as belonging to the boundary of $\Gamma$. A
schematization of this procedure is depicted in figure
\ref{changes}a). By changing the initial point along the
considered segment of the bisectrix, we are able to define the
whole boundary of $\Gamma$ above the bisectrix with a good degree
of precision. Then, the symmetry properties of $\Gamma$ allow us
to easily deduce the portion of its boundary lying below the
bisectrix. In figure \ref{Stabilityfigure} we present the estimate
for the the boundary of the bistable region $\Gamma$ obtained
following this strategy.

We conclude that the boundary of $\Gamma$ is constituted by the
bifurcation points of the system. We note that in the more general
case of the 2D $(\Phi_S,\Phi_N)$ plane,
$(\Phi_S=\Phi_{inf},\Phi_N=\Phi_{inf})$ and
$(\Phi_S=\Phi_{sup},\Phi_N=\Phi_{sup})$ result to be cusp points
\citep{Dijkstra2001}.

%%%%%%%%%%%%%%%%%%%%%%%%%%%%%%%%%%%%%%%%%%%%%
%%%%%%%%%%%%%%%%%%%%%%%%%%%%%%%%%%%%%%%%%%%%%
\section{Effects of transient perturbations}
\label{generalstability}
%%%%%%%%%%%%%%%%%%%%%%%%%%%%%%%%%%%%%%%%%%%%%
%%%%%%%%%%%%%%%%%%%%%%%%%%%%%%%%%%%%%%%%%%%%%

The analysis we have performed captures the equilibrium properties
of the system, but is not sufficient to gain insight on the
response of the system to transient changes in the forcings, which
in general can range from instantaneous to quasi-static
perturbations. Such sort of problems has been first investigated
in the seminal paper by \cite{StockerandSchmittner1997} with the
purpose of determining how efficiently the negative feedbacks of
the system can counteract external perturbations, depending on the
temporal patters of the destabilizing forcings.

In our case, it is reasonable to expect that, if we change
$\Phi_N$ at a finite rate, the system can destabilize before
reaching the boundary of $\Gamma$. In fact, a fast perturbation
can overcome the ability of the advective feedback to stabilize
the system. We also expect that the faster the perturbation, the
stronger such effect, \textit{i.e.} the smaller the total
perturbation required to obtain destabilization.

As in the previous case, the analysis starts by considering as
initial equilibria the northern sinking states under symmetric
forcing. In this case, we increase the value of $\Phi_N$ at a
constant rate $r_f$ over a time $t_0$, we let the system adjust
for a time $10 \tau$, so that transients can die out, and then
decrease $\Phi_N$ back to the initial value at the slow rate
$-r_s$, corresponding to a quasi-static change. We schematically
depict such strategy for three different values of $r_f$ in figure
\ref{changes}b). If at the end of the process the initial state is
not recovered and southern sinking state is instead realized, the
system has made a transition to the other branch of the multiple
equilibria. Depending on the choice of $r_f$, we obtain different
values for the critical perturbation causing such transition.
Along the lines of the quasi-static analysis, varying the initial
point, we can obtain for each value of $r_f$ a curve which
describes the critical perturbations.

\begin{figure}
 \centering
\includegraphics[angle=270, width=1.0\textwidth]{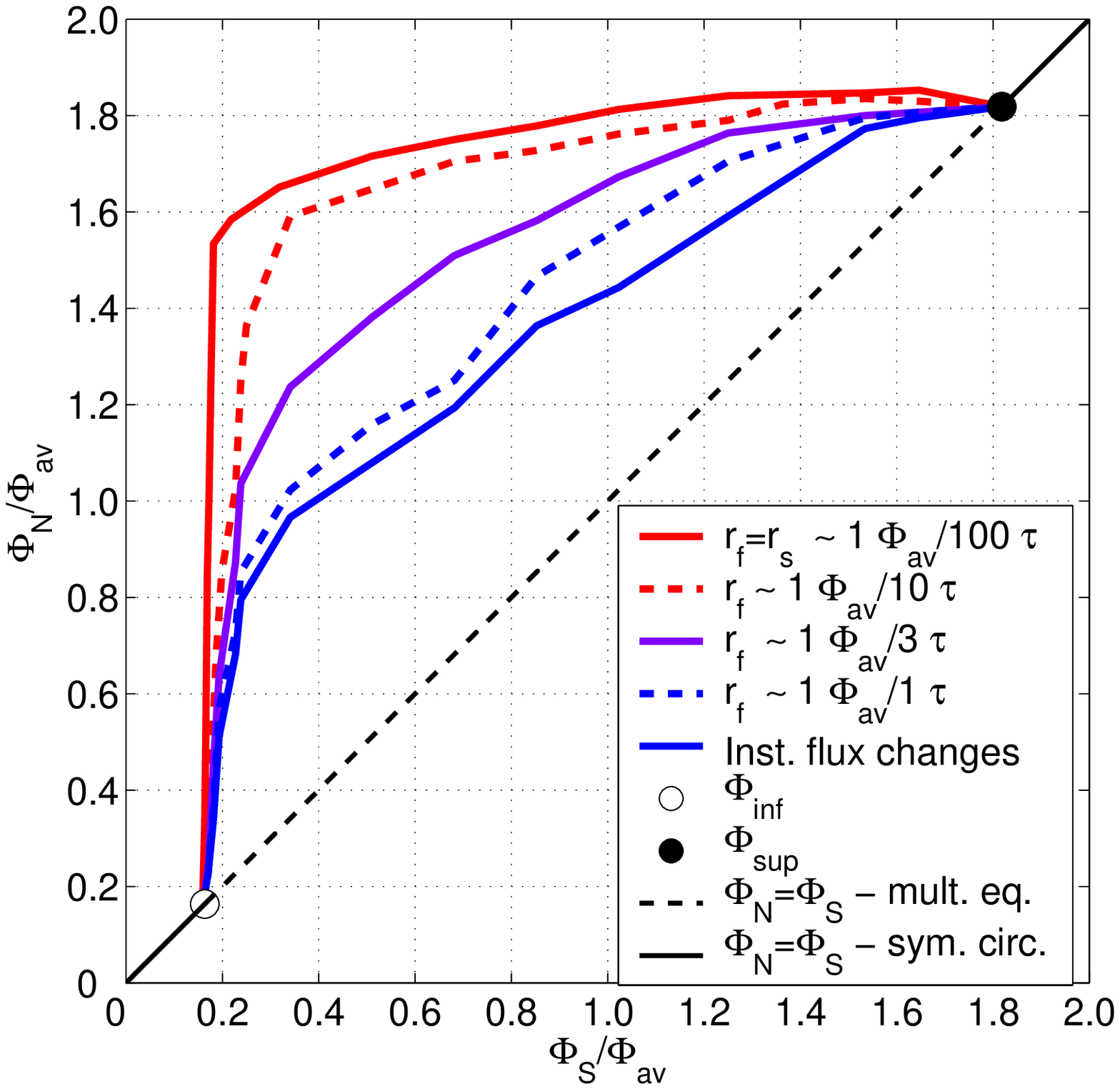}\\
\caption{Critical forcings for the collapse of the northern
sinking pattern in the space $(\Phi_S,\Phi_N)$. Various temporal
patterns of the forcings are considered.} \label{Stabilityfigure2}
\end{figure}

In figure \ref{Stabilityfigure2} we report the curves obtained by
selecting, from fast to slow, $r_f=\infty$, $r_f=\Phi_{av}/\tau$,
$r_f=\Phi_{av}/3 \tau$, $r_f=\Phi_{av}/10\tau$, and
$r_f=r_s=\Phi_{av}/100\tau$.

\begin{figure}
 \centering
\includegraphics[angle=270, width=1.0\textwidth]{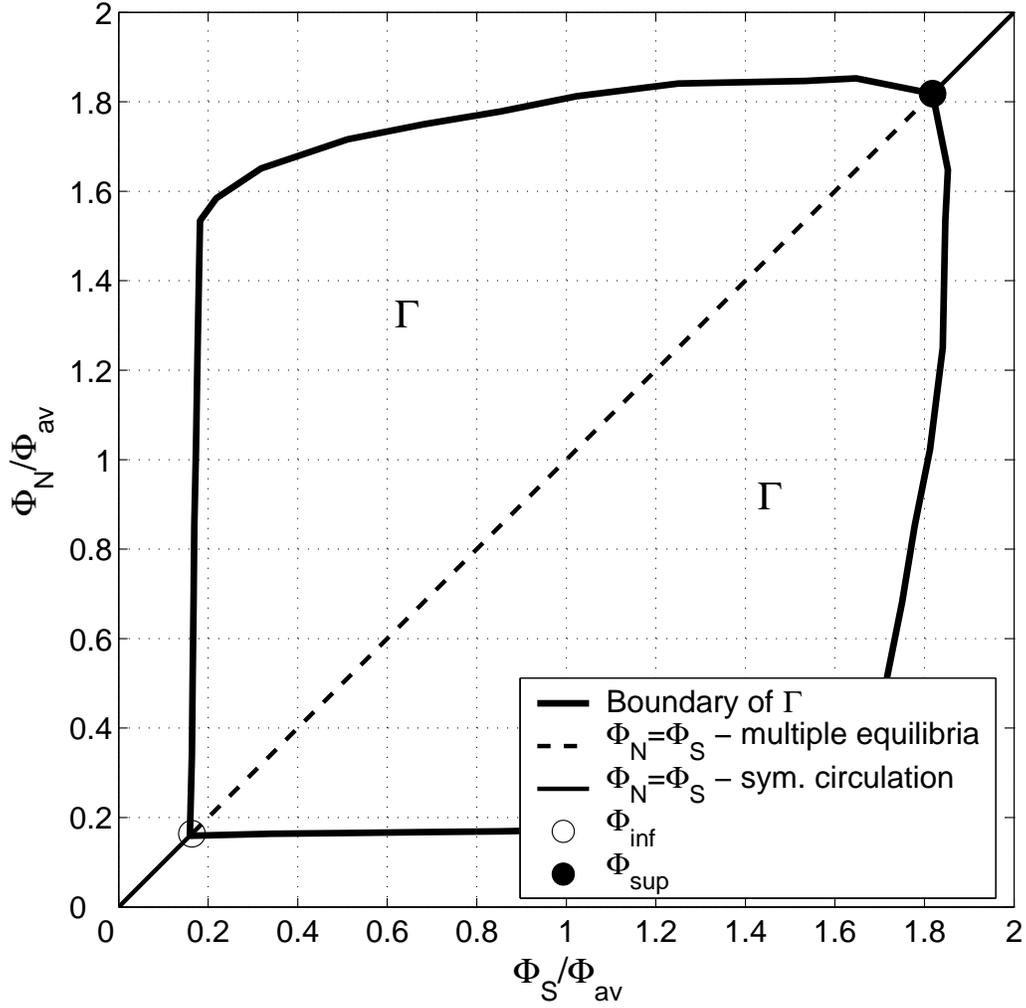}\\
\caption{Stability graph of the system in the space
$(\Phi_S,\Phi_N)$. The thick black line delimitates the
bistability region $\Gamma$. Along the diagonal, solid lines
represent the bistable states having antisymmetric circulation
patters, while the dashed line represent represent the stable
symmetric circulation patters.} \label{Stabilityfigure}
\end{figure}

If we select $r_f=r_s$, we obtain by definition the previously
described upper branch of the boundary of $\Gamma$ depicted in
figure \ref{Stabilityfigure}, since the presence of the relaxing
time $ 10 \tau$ is not relevant.

On the other extreme, if we apply instantaneous changes of
$\Phi_N$ ($r_f=\infty$), we obtain information on the minimum
change in $\Phi_N$ that is needed to destabilize the system for
any initial state having symmetric surface forcing. In fact, the
corresponding curve is the closest to the bisectrix.

Considering intermediate values of $r_f$, we obtain consistently
that the curves of the critical perturbations lye within the two
extremes obtained with $r_f=r_s$ and $r_f=\infty$. Moreover, we
have that the curves are properly ordered with respect to the
value of $r_f$, \textit{i.e.} the smaller $r_f$, the closer the
corresponding curve to the upper branch of the boundary of
$\Gamma$.

Previous studies, albeit performed with coupled EMICs, obtain a
qualitatively similar dependence of thresholds on the rate of
increase of the forcings \citep{StockerandSchmittner1997,
SchmittnerandStocker1999}, while in other studies (on GCMs) where
the full collapse of the THC is not obtained, it is nevertheless
observed that the higher the rate of increase of the forcing, the
larger the decrease of THC realized \citep{StoufferandManabe1999}.

The most important result is that we can identify two separate
regimes. If the surface forcing changes with a rate faster than
$\Phi_{av}/\tau$, the response of the system is virtually
identical to the case of instantaneous changes. On the other side,
if the rate of change is smaller that $\Phi_{av}/10 \tau$, the
response of the system is very close to the case of quasi-static
perturbations. The curve corresponding to $r_f=\Phi_{av}/3 \tau$
is geometrically about midway between these two regimes and is not
patched to either. We have that the response of the system to
varying external perturbations dramatically changes when the time
scale of the variation of the external forcing changes by only one
order of magnitude. Therefore, we can interpret $r_c=\Phi_{av}/3
\tau$ as an estimate of the critical rate of change of the
hydrology of the system.

It follows that with $r_c$ we identify a relation between the
class of changes in the external forcing that distinctively affect
the stability of the system and the internal time scale of the
system.

\begin{figure}
 \centering
\includegraphics[angle=270, width=1.0\textwidth]{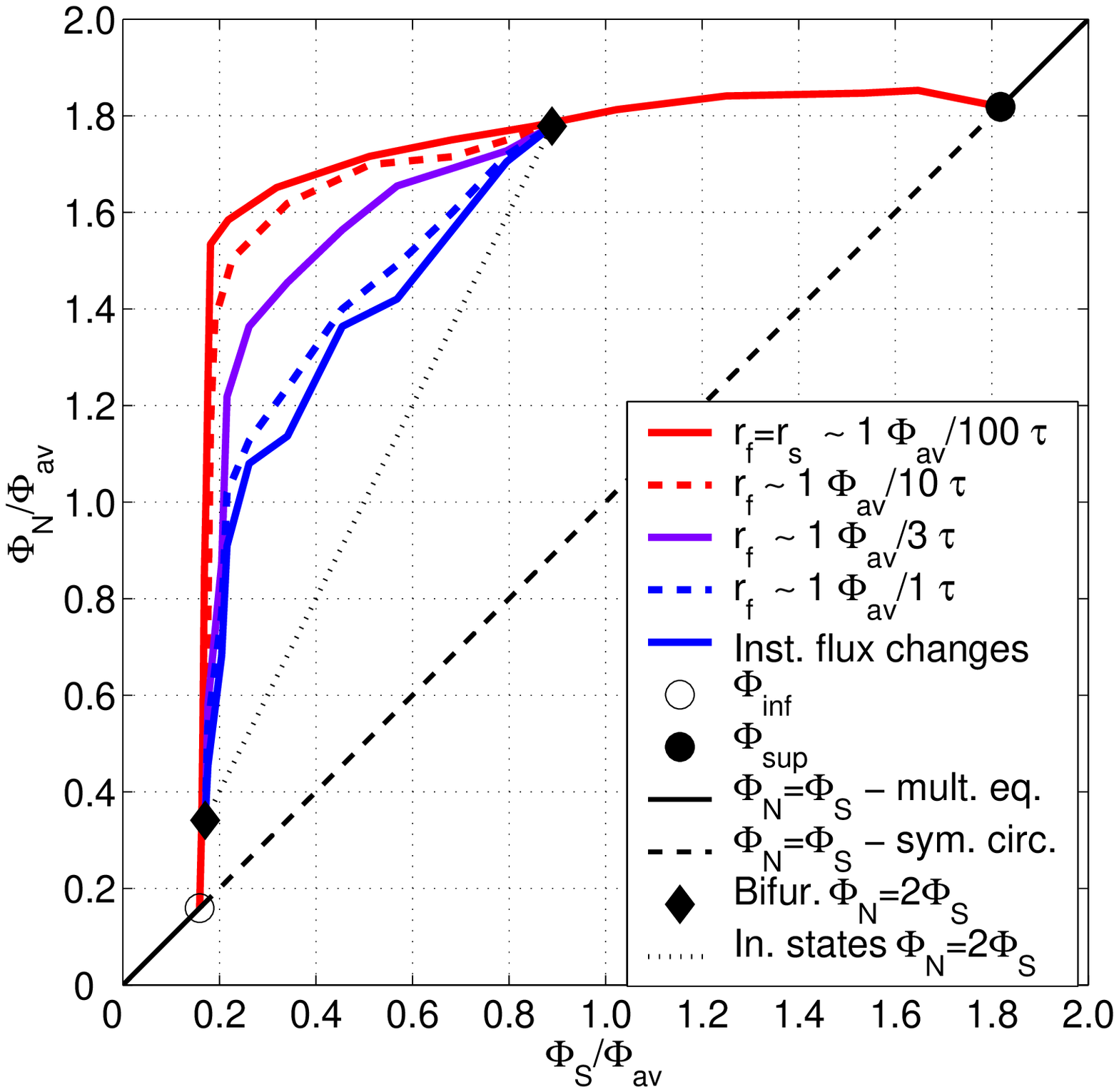}\\
\caption{Critical forcings for the collapse of the northern
sinking pattern in the space $(\Phi_S,\Phi_N)$. Various temporal
patterns of the forcings are considered.} \label{Stabilityfigure3}
\end{figure}

The previous results prove to be robust with respect to changes in
the freshwater forcing of the initial states. As an example, in
figure \ref{Stabilityfigure3} we show the results of a similar
analysis referring to initial states having non-symmetric surface
freshwater forcing ($\Phi_N=2\Phi_S$).

We explored the behavior of the system in the $(\Phi_S,\Phi_N)$
plane considering only changes in $\Phi_N$ for computational
convenience. Nevertheless, coherent results can be obtained by
changing both parameters. In this case $r_f$ has to be interpreted
as the sum of the absolute values of the rates of change of the
two parameters.

Changing the parameter $\Delta \overline{T}$ implies a change in
the values of $\Phi_{inf}$ and $\Phi_{sup}$, since the meridional
gradient of the total buoyancy forcing is changed. However, the
response of the THC to transient perturbations does not change
qualitatively. Instead, figures
\ref{Stabilityfigure}-\ref{Stabilityfigure3} must be rescaled
linearly with the proper values of $\Phi_{sup}$ and $\Phi_{inf}$.
Such linear relation is a direct consequence of the use of a
linearized equation of state for the sea water.

%%%%%%%%%%%%%%%%%%%%%%%%%%%%%%%%%%%%
%%%%%%%%%%%%%%%%%%%%%%%%%%%%%%%%%%%%
\section{Conclusions}
\label{concl}
%%%%%%%%%%%%%%%%%%%%%%%%%%%%%%%%%%%%
%%%%%%%%%%%%%%%%%%%%%%%%%%%%%%%%%%%%

This work provides a complete analysis of the stability of the
ocean system under examination with respect to perturbation to the
hydrological cycle. We have provided a simple description of the
profile of the net freshwater flux into the ocean which is fully
specified when only two parameters, which are related to the total
freshwater budget of the two high-latitudes regions, are
specified.

We have first analyzed the bifurcations of the symmetric system,
which might be taken as the prototype of a system that has equal
probabilities of falling in two different equilibrium
configurations. We have found that the system is characterized by
two bifurcations, which delimitate a domain of multiple
equilibria.

We have then extended the study to the general case where
asymmetries in the hydrology are considered. We have produced a
two-dimensional stability graph and have pointed out the presence
of a region $\Gamma$ where multiple equilibria are realized. Our
results summarize the information that can be obtained with
multiple hysteresis studies.

% and propose a new methodology for the
%study of the THC system.

%In the real ocean, the three
%dimensional dynamics implies the existence of different feedback
%mechanisms, other than the advective feedback which is at the core
%of  the instabilities seen in this work.
%In all cases, the
%the effect of transient changes must be evaluated

%By looking at the dynamics of hysteresis cycles of asymmetric
%(pole to pole circulation) equilibria of the thermohaline
%circulation, we are addressing some fundamental questions
%concerning the non-linear behavior of a simple advective system.

%We have observed that changing the resolution within a relatively
%wide range ($48 \times 12$-$120\times 40$) does not appreciably
%alter the position of the bifurcation points.
%We go a step forward than \cite{Dijkstra1999} by looking,
%systematically although not exhaustively, at the temporal
%evolution of finite amplitude modifications of the hydrological
%cycle that are able to destabilize the equilibria of a simple
%advective-convective system.
%In the real ocean, the three
%dimensional dynamics implies the existence of different feedback
%mechanisms, other than the advective feedback which is at the core
%of every instability process which is seen in this work
%\citep{LenderinkandHaarsma1994}.

In this study we emphasize that  the rate at which a perturbation
in the hydrological cycle is applied to a simple model of the THC
may dramatically affect its stability. When general time-depends
perturbation to the hydrological cycle are considered, we obtain
that the shorter the time scale of the forcing, the smaller the
total perturbation required to disrupt the initial pattern of the
circulation. The observed relevance of the temporal scale of the
forcing in determining the response of our system to perturbations
affecting the stability of the THC agrees with the findings of
\cite{TzipermanandGildor2002}, \cite{
Lucarini2003aa,Lucarini2003bb} for box models,  of
\cite{StockerandSchmittner1997} and
\cite{SchmittnerandStocker1999} in the context of EMICSs, and of
\citep{ManabeandStouffer1999a}, \cite{ManabeandStouffer1999b},
\cite{ManabeandStouffer2000}, and \cite{StoufferandManabe1999} in
the context of GCMs. Moreover, the saturation and patching effect
observed for slowly and rapidly increasing perturbations, which
allows the definition of \textit{slow} and \textit{fast} regimes,
respectively, agrees qualitatively with the findings of
\cite{Lucarini2003aa,Lucarini2003bb} for box models, and resembles
some of the results - albeit obtained with a coupled model and
considering CO$_2$ increases - presented in
\cite{StockerandSchmittner1997} and in
\cite{SchmittnerandStocker1999}.

The main conceptual improvement we propose in this work is the
existence of a relation between the critical rate of change of the
forcing and  the characteristic advective time scale of the
system.

We notice that the advective time scale results to depend on the
inverse of the square root of $K_V$, as discussed in appendix A.
Therefore, a very important consequence of this analysis is that
the efficiency of the vertical mixing might be also one of the key
factors determining the response of the THC system to transient
changes in the surface forcings. Future work should specifically
address the details of the functional dependence of the critical
rate on the the vertical diffusivity.

Other relevant improvements to the present study could be the
adoption of a more complex ocean model, descriptive of other ocean
basins, as well as the consideration of a simplified coupled
atmosphere-ocean model, where the effects of the radiative forcing
can be more properly represented.

%In particular, Moreover, our results  confirm qualitatively the
%findings of a study performed on a simple 3-box model
%\citep{Lucarini2003aa}. In quantitative terms, the results
%obtained in this study with a two-dimensional model differ from
%those of the 3-box model, where a very large difference was found
%between the response of the system to fast and slow perturbation,
%where fast and slow was determined by the only relevant time scale
%of the system. The consideration of a larger number of degrees of
%freedom blurs the distinction between slow and fast perturbations.
%The active degree of freedom which is gained in this study is the
%equatorial region. As a consequence, the system has only a limited
%domain of bistability, in contrast with the findings of
%\cite{Scottetal.1999}. In such analysis they show that in the
%$(\Phi_S,\Phi_N)$ plane it is possible to find an unbounded region
%of multiple equilibria.

%Future work will include understanding the sensitivity of the
%stability diagram on the the asymmetries in the $\tau$ profile.

\subsection*{Acknowledgment}
We wish to thank for technical and scientific help Fabio Dalan,
Antonello Provenzale, Peter H. Stone, and Antonio Speranza.
% ------------------------------------------------------------------------
%GATHER{Xbib.bib}   % For Gather Purpose Only
%GATHER{Paper.bbl}  % For Gather Purpose Only
\clearpage
\newpage

\section*{Appendix A. Relevance of the vertical diffusivity $K_V$}

\begin{figure}
 \centering
\includegraphics[angle=270, width=1.0\textwidth]{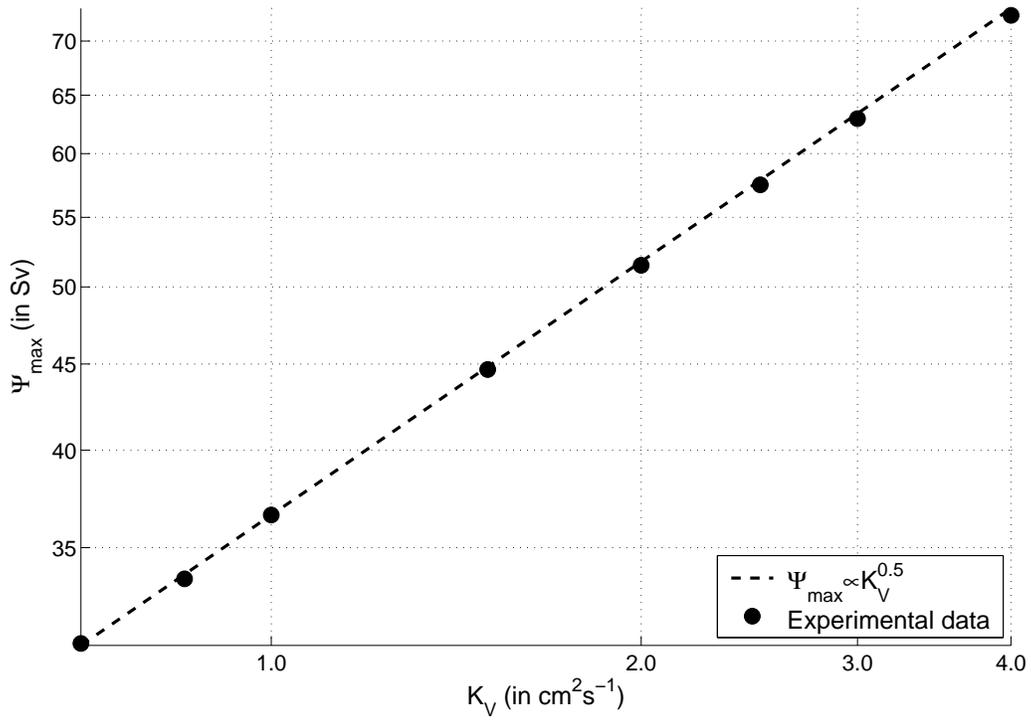}\\
\caption{Dependence of maximum value of the THC on the value of
the vertical diffusivity for $\overline{T} = 15.0 ^{\circ}C$,
$\Delta \overline{T} = 23.5^{\circ}C$, and
$\Phi_N=\Phi_S=\Phi_{av}$} \label{KVfigure}
\end{figure}

The vertical diffusivity $K_V$, or, equivalently, the diapycnal
diffusivity $K_D$, is the critical parameter controlling the
maximum THC strength $\Psi_{max}$ in ocean models
\citep{Bryan1987,Wright1992}. On the other hand, an estimate of
its value in the real ocean is a  subject of current research
\citep{Gregg2003}. Scaling theories proposing a balance between
vertical diffusion and advection processes suggest, in the case of
three-dimensional hemispheric model of the Atlantic ocean, a power
law dependence $\Psi_{max} \sim K_V ^{2/3}$
\citep{Zhang1999,Dalan2004}.
%Studies have found different power laws when full interhemispheric
%of global models are considered, and some theoretical explanations
%for the discrepancy have been proposed (Dalan, F. and P. H. Stone,
%personal communication, 2004).
In the case of two-dimensional models, the expected dependence is
$\Psi_{max} \sim K_V ^{1/2}$ \citep{Knutti2000}. This relation is
verified in our model in the range $0.6 cm^2 s^{-1} < K_V < 4.0
cm^2 s^{-1}$ (figure \ref{KVfigure}). In this study, the value of
$K_V$ has been selected so that the corresponding northern sinking
equilibrium state characterized by an hydrological cycle
determined by $\Phi_S=\Phi_N=\Phi_{av}$ has an overturning
circulation  $\sim 30 Sv$. With this choice of $K_V$ we can define
a characteristic time scale $\tau$ for the system as
$\tau=V/\Psi_{max} \sim 350 y$. Similarly, if we change $K_V$ over
the range shown in figure \ref{KVfigure}, the advective time-scale
would range over the interval $ 150 y < \tau < 400 y$. These
considerations will be useful in the final discussion of our
results. Given the parameters chosen for our simulations, our
model integrations estimate $\Phi_{inf} \sim 0.04 Sv$ and
$\Phi_{sup} \sim 0.73 Sv$, so that $\Phi_{av}$ results to be $\sim
0.39 Sv$.

\newpage

\newpage

%\newpage
%\begin{figure} \centering
%\includegraphics[angle=270, width=1.0\textwidth]{StabilityTlarge}\\
%\caption{Bifurcation diagram of the system under consideration ;
%x-axis: intensity of the hydrological cycle $\Phi_N=\Phi_S$;
%y-axis: average vorticity $\Psi_{av}$. Solid lines: stable
%equilibria; dashed line: unstable equilibria. Bifurcation points
%are indicated.}\label{BIFfigure}
%\end{figure}
%\newpage
%\begin{figure}
% \centering
%\includegraphics[angle=270, width=1.0\textwidth]{Hysteresis9090}\\
%\caption{Hysteresis graph obtained varying quasi-statically
%$\Phi_N$ with
%$\Phi_N\left(0\right)=\Phi_S=1/2\times(\Phi_{inf}+\Phi_{sup})$.
%$\Phi_{N}^{lo}\left(\Phi_N(0)\right)$ and
%$\Phi_{N}^{up}\left(\Phi_N(0)\right)$ are indicated.}
%\label{Hysteresisfigure}
%\end{figure}
%\newpage
%\begin{figure}
% \centering
%\includegraphics[angle=270, width=1.0\textwidth]{Hysteresis9090counter}\\
%\caption{Hysteresis graph obtained varying quasi-statically
%$\Phi_S$ with
%$\Phi_S\left(0\right)=\Phi_N=1/2\times(\Phi_{inf}+\Phi_{sup})$.
%$\Phi_{S}^{lo}\left(\Phi_S(0)\right)$ and
%$\Phi_{S}^{up}\left(\Phi_S(0)\right)$ are indicated.}
%\label{Hysteresisfigure2}
%\end{figure}
%\newpage

\newpage

\newpage

\newpage

\newpage

%\newpage
%\begin{figure} \centering
%\includegraphics[angle=270, width=1.0\textwidth]{StabilityTsmall}\\
%\caption{Bifurcation diagram of the system under consideration
%with $\Delta \tau=15^{\circ}C$; x-axis: intensity of the
%hydrological cycle $\Phi_N=\Phi_S$; y-axis: average vorticity
%$\Psi_{av}$. Solid lines: stable equilibria; dashed line: unstable
%equilibria. Bifurcation points are indicated.}\label{BIFfigure2}
%\end{figure}
%\newpage
%\begin{figure}
% \centering
%\includegraphics[angle=270, width=1.0\textwidth]{bistabilitysmall}\\
%\caption{Stability graph of the system in the space
%$(\Phi_S,\Phi_N)$ with $\Delta \tau=15^{\circ}C$. The red line
%delimitates the bistability region $\Gamma$. The dots represent
%the experimental points. Along the diagonal, solid lines represent
%the bistable states having antisymmetric circulation patters,
%while the dashed line represent represent the stable symmetric
%circulation patters.} \label{Stabilityfigure3}
%\end{figure}

\newpage

\end{document}